\documentclass{ifacconf}
\usepackage{graphicx}      % include this line if your document contains figures
\usepackage{natbib}        % required for bibliography
\usepackage{tabularx}
\usepackage{marginnote}
\usepackage{amsmath}
\usepackage[table]{xcolor}% http://ctan.org/pkg/xcolor
\usepackage{multirow}
\usepackage{amsfonts,bm} % Math packages
\usepackage{framed} % Framing content 
\usepackage{nomencl}
\usepackage{float}
\usepackage{optidef}
\usepackage{mathtools}
\DeclareMathSymbol{\sm}{\mathbin}{AMSa}{"39} %short minus
\newcommand{\ind}[1]{{_{\mathrm{#1}}}} %  Alex
\usepackage{background}
\backgroundsetup{contents = This paper accepted for oral presentation at 10th Annual IFAC Advances in Automotive Control Symposium, opacity = 0.45, scale = 1, color = blue, angle = 0, vshift = -12cm}

\newcommand{\nox}{$\mathrm{NO_{x}}$}

\nomenclature{ML}{Machine Learning}
\nomenclature{MPC}{Model Predictive Control}
\nomenclature{LSTM}{Long-short term memory}
\nomenclature{ICE}{Internal Combustion Engine}
\nomenclature{LPV}{linear parameter-varying}
\nomenclature{ECU}{Electronic control unit}
\nomenclature{NMPC}{Nonlinear Model Predictive Control}
\nomenclature{LMPC}{Linear Model Predictive Control}
\nomenclature{ARX}{Autoregressive with Extra Input}
\nomenclature{SOI}{Start Of Injection}
\nomenclature{FQ}{Fuel Quantity}
\nomenclature{DPM}{Detailed Physical Model}
\nomenclature{SVM}{Support Vector Machine}
\nomenclature{NOx}{Nitrogen Oxide}
\nomenclature{RNN}{Recurrent Neural Network}
\nomenclature{VGT}{Variable-Geometry Turbocharger }
\nomenclature{OCP}{Optimal Control Problem}
\nomenclature{SQP}{Sequential Quadratic Programming}
\nomenclature{ESM}{Engine Simulation Model}

\begin{document}
\begin{frontmatter}

\title{Machine Learning Integrated with Model Predictive Control for Imitative Optimal Control of Compression Ignition Engines}

\thanks{Corresponding Author: Armin Norouzi-(e-mail: norouziy@ualberta.ca). }

% \author[First]{Armin Norouzi}
% \author[First]{Saeid Shahpouri}
% \author[First]{Mahdi Shahbakhti}
% \author[First]{Charles Robert Koch}
% \address[First]{Mechanical Engineering Department,~University~of~Alberta, Edmonton, Canada}

\author[1]{Armin Norouzi} %113371
\author[1]{Saeid Shahpouri} %137059
\author[1]{David Gordon} %129281
\author[2]{Alexander Winkler} %135760
\author[3]{Eugen Nuss} %147980
\author[3]{Dirk Abel} 
\author[2]{Jakob Andert}
\author[1]{Mahdi Shahbakhti} %30396
\author[1]{Charles Robert Koch} %19775
\address[1]{Department of Mechanical Engineering, University of Alberta, Edmonton, Canada}
\address[2]{Teaching and Research Area Mechatronics in Mobile Propulsion, RWTH Aachen University, Aachen, Germany}
\address[3]{Institute of Automatic Control, RWTH Aachen University, Germany}

\begin{abstract}
The high thermal efficiency and reliability of the compression-ignition engine makes it the first choice for many applications. For this to continue, a reduction of the pollutant emissions is needed. One solution is the use of Machine Learning (ML) and Model Predictive Control (MPC) to minimize emissions and fuel consumption, without adding substantial computational cost to the engine controller. ML is developed in this paper for both modeling engine performance and emissions and for imitating the behaviour of a Linear Parameter Varying (LPV) MPC. Using a support vector machine-based linear parameter varying model of the engine performance and emissions, a model predictive controller is implemented for a 4.5~L Cummins diesel engine. This online optimized MPC solution offers advantages in minimizing the \nox~emissions and fuel consumption compared to the baseline feedforward production controller. To reduce the computational cost of this MPC, a deep learning scheme is designed to mimic the behavior of the developed controller. The performance in reducing NOx emissions at a constant load by the imitative controller is similar to that of the online optimized MPC, however, the imitative controller requires 50 times less computation time when compared to that of the online MPC optimization.  
\end{abstract}

\begin{keyword}
Diesel engines, Linear Parameter Variable Model, Machine learning, Support Vector Machine, Model Predictive Control, Deep Learning
\end{keyword}
\end{frontmatter}
%===============================================================================

%%%%%%%%%%%%%%%%%%%%%%%%%%%%%%%%%%%%%%%%%%%
\section{Introduction}

Compression Ignition (CI) engines are commonly used in various transportation applications, from public transportation to personal vehicles. The diesel engine has a high thermal efficiency, long lifetime as well as fuel economy advantages at full-load and part-load conditions when compared to spark ignition engines~\citep{ortner2007predictive}. Although diesel engines have many advantages and are commonly used, they play a significant role in the environmental pollution problems worldwide~\citep{recsitouglu2015pollutant}. Hybridization and electrification are proceeding rapidly for passenger vehicles, but the uptake has been slow for commercial heavy-duty trucks due to the high battery costs and decreased cargo capacity~\citep{heid2017s}.

Real Driving Emissions (RDE) requirements have been implemented beginning with Euro 6d. The challenge with RDE requirements are that test results are significantly affected by various external factors including the ambient conditions, traffic, and driver behavior. Based on RDE legislation, engines should operate cleanly under all conditions, which makes engine design and calibration much more challenging~\citep{mpcreviean}. Complying with RDE legislation is a significant shift from the previous legislation. Intelligent engine control strategies that take advantage of Machine Learning~(ML) methods and utilize model-based control are one way to make substantial progress toward meeting the rigorous emission regulations.

Model Predictive Control (MPC) has been shown to be an effective model based control strategy. This is the result of MPC considering constraints on inputs, outputs, and states. Additionally, MPC offers optimal performance control by utilizing a future horizon while optimizing the current control law~\citep{ICEMPC15, stewart2008model, WINKLER2021359}. MPC has been used for CI control of various different engines from light to heavy duty and for broad range of applications~\citep{ICEMPC15, ICEMPC1, yin2020model, liu2021simultaneous}. Despite the many benefits of MPC, it has a high computational cost and the performance depends on the accuracy of the embedded model. There is a trade-off based on the complexity of the model between prediction accuracy and computation time. This is the main reason that physics-based based models, such as detailed 3D combustion models, which offer a high prediction accuracy, have seen limited MPC implementation due to high computational times. These models first need to be simplified or linearized to be feasible for MPC implementation~\citep{stewart2008model, norouzi2019integral}.

A state-space Linear Parameter Varying (LPV) model is capable of providing an accurate combustion model while utilizing a simpler structure~\citep{ICEMPC7}. LPV is a state-space system that updates the system matrices based on scheduling parameters. The LPV model can fill the gap between complex nonlinear models and inaccurate but simple linear models. It employs an array of linear models characterized by scheduling variables. The LPV model has been used in internal combustion control within an MPC framework in literature for both CI and Spark Ignition (SI) engines~\citep{ICEMPC7, ICEMPC8}.

The traditional use of an LPV model requires the evaluation of a complex model in a grid of scheduling parameters; however, ML provides a systematic way to create a state-space LPV model directly from measurement data. SVM-LPV~\citep{rizvi2015iv} uses a Least-Square Support Vector Machine (LS-SVM) framework to update state-space matrices. This method has been previously applied to Internal Combustion (IC) engine performance modeling~\citep{ICEMPC7, ICEMPC8}. In this paper, a Support Vector Machine-based state-space LPV algorithm (SVM-LPV) is used to model engine performance and the engine-out \nox~ emissions to show the capability of the SVM-LPV technique. To the best of authors’ knowledge, this paper presents the first study to use this technique for emission modeling and control of an IC engine. To make this algorithm reach the optimum solution, a Beysian optimization has been added to optimize the hyperparamters of the SVM. 

Modeling of the engine is the first use of ML in this paper. Then cloning the behaviour of an MPC is the second use of ML presented in this paper which is called learning, approximate, or imitation MPC. This learning MPC utilizes a deep neural network to imitate the behaviour of the designed MPC and it has been used successfully in vehicle dynamics ~\citep{zhang2019safe}, and the heating, ventilation, and air conditioning (HVAC) industry \citep{toub2019mpc}. In this case is has been shown to provide a significant computational time reduction in comparison with traditional online optimization of MPC. However, to the authors' knowledge this imitative LPV-MPC has not been used for IC engine control.

Based on the current literature, this paper has the following contributions:

\begin{enumerate}
    \item Machine Learning based modeling for MPC design
    \begin{enumerate}
    \item An SVM-LPV model is adapted to develop a linear parameter-varying model for engine-out \nox~ emissions and engine performance metrics
    \item An LPV MPC based on a ML model to minimize engine-out emissions and fuel consumption while maintaining the same output torque performance and comparison with a benchmark model (Cummins calibrated ECU-based GT-power model)
\end{enumerate}
\item Machine Learning based control to reduce computational time of MPC
    \begin{enumerate}
    \item Developed imitation based controller using deep neural network to clone behavior of LPV-MPC to reduce the computational time of optimization
\end{enumerate}
\end{enumerate}

%%%%%%%%%%%%%%%%%%%%%%%%%%%%%%%%%%%%%%%%%%%
\section{Engine Simulation Model}

A 4.5-liter medium-duty Cummins Diesel engine is the focus of this work. The Cummins QSB4.5 160 Diesel engine is a turbocharged inter-cooled engine meeting Tier 3 emissions legislation. A dSPACE MicroAutoBox II is used to control the intake manifold pressure, engine speed, load, injected fuel amount, and fuel rail pressure. A Kistler piezoelectric pressure sensor is used to measure the in-cylinder pressure. Engine-out Nitrogen Oxides (\nox) emissions from the engine are measured using a Bosch sensor with ECM electronics (P/N: 06-05). Further experimental setup details can be found in~\citep{norouzi2020correlation, 2020CCTASVMDiesel}. The engine is used to parameterize the simulation model.

A Detailed Physical Model (DPM) of the real engine was created using GT-Power to test the proposed control strategies in an Engine Simulation Model (ESM) platform. This model contains several chemical and physical sub-models that are used to simulate the complex combustion and gas exchange processes. This experimentally validated model is from our previous studies with an accuracy of $\pm$5.8\%, $\pm$4.6\%, and $\pm$18.1\% in the prediction of the maximum in-cylinder pressure, intake manifold pressure, and \nox, respectively~\citep{saeed2021MECC, shahpournyenergies}.

%%%%%%%%%%%%%%%%%%%%%%%%%%%%%%%%%%%%%%%%%%%
\section{Methodology}
\begin{figure}
    \centering
    \includegraphics[trim = 0 0 0 0, clip, width = 0.45\textwidth]{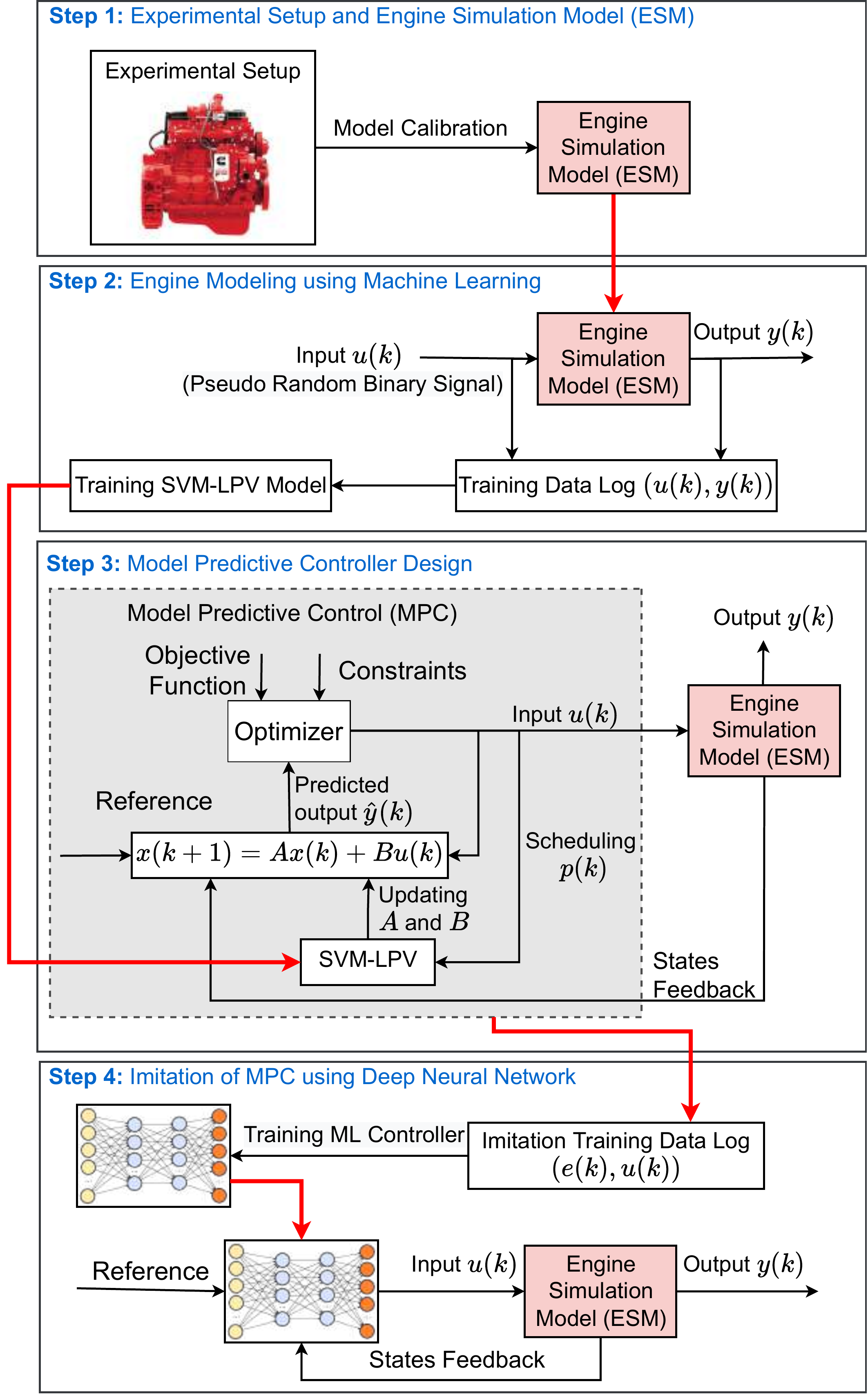}
    \caption{Modeling and controller design procedure based on Engine Simulation Model (ESM)}
    \label{fig:offlinemodeling}
\end{figure}

In this paper, two methods of combining ML and MPC are presented consisting of ML-based modeling and ML imitation control which tested in simulation. An overview of the methodology is depicted in Fig.~\ref{fig:offlinemodeling}. First, randomly generated inputs are fed into the ESM and the output engine performance is recorded for control modeling. A Least-square Support Vector Machine based Linear Parameter-Varying (SVM-LPV) model is then developed using the input-output data. This model is used for the design of the LPV-MPC controller. Finally, this MPC controller is used to train the ML based imitation controllers. To assess, the LPV-MPC controller performance it is compared to a Linear Autoregressive with Extra Input (ARX) based linear MPC using a GT-power/MATLAB/SIMULINK co-simulation.

\section{Modeling}
\subsection{Support Vector Machine based Linear Parameter Varying (LPV) Model}

The LPV model is defined as:
\begin{equation}
\begin{split}
    \mathbf{x}({k+1}) &= A\left(p(k)\right) \mathbf{x}(k) +  B\left(p(k)\right) \mathbf{u}(k)\\ \label{eq:lpv_1}
    \mathbf{y}(k) &=  C\left(p(k)\right) \mathbf{x}(p(k)) +  D\left(p(k)\right) \mathbf{u}(k) 
\end{split}
\end{equation}

Where the state matrices (A,B,C and D) are a function of scheduling parameters, \( p(k)\). An SVM-based algorithm is used to update the state matrices. In this study \( u(k)\). \( x(k)\), and \( y(k)\) are defined as
\begin{align}
\begin{split}
	u(k) &= \begin{bmatrix} \text{FQ}(k) & \text{SOI}(k) & \text{VGT}(k) \end{bmatrix}^T, \\
	 x(k) &= \begin{bmatrix} T_\text{out}(k) & P_\text{man}(k) & \text{NO}_x(k) \end{bmatrix}^T, 
    \\
	 y(k) &= \begin{bmatrix} T_\text{out}(k) & \text{NO}_x(k) \end{bmatrix}^T, 
    \\
\end{split}\label{eq:statescontrols}
\end{align}
where $FQ(k)$ is Fuel Quantity (FQ), $SOI(k)$ is the Start Of main Injection (SOI), $VGT(k)$ is the Variable Geometric Turbine (VGT) rate, $T_{\text{out}}$ is the engine output torque, $P_{\text{man}}$ is the intake manifold pressure, and $\text{NO}_x(k)$ is the engine-out Nitrogen Oxides (\nox) emissions. The SVM-LPV algorithm developed in \citep{rizvi2015iv} is adapted for this specific problem. Additionally, to tune the hyperparameters of the SVM-LPV, a Bayesian optimization is implemented. It is assumed that the output of the model is equal or partially equal to the states of the system and that the system states are measurable. Thus, the matrix \( C \) is not scheduled, and matrix \( D \) is identically zero. Then, the model can be simplified as
\begin{equation}
\begin{split}
    \mathbf{x}({k+1}) &= A\left(p(k)\right) \mathbf{x}(k) + B\left(p(k)\right) \mathbf{u}(k)
    \label{eq:lpv_2}
\end{split}
\end{equation}
where $u(k) \in \mathbb{R}^{n_u} $ , $x(k) \in \mathbb{R}^{n_x} $, and $p(k) \in \mathbb{R}^{n_p} $ are inputs, states, and scheduling parameter at $k$, and $A  \in \mathbb{R}^{(n_x \times n_x)} $ and $B  \in \mathbb{R}^{(n_x \times n_u)} $ are state-space model matrices ($n_x$ and $n_u$ are the number of states and manipulated variables, respectively). To formulate our problem in an SVM framework, \(A(p(k))\) and \(B(p(k))\) can be written as
\begin{equation}\label{eq:lpv_3}
\begin{split}
    A\left(p(k)\right)=& W_1 \phi_1\left(p(k)\right),~~~~  
    B\left(p(k)\right) = W_2 \phi_2\left(p(k)\right) \\
\end{split}
\end{equation}
where \( \phi_1 \in \mathbb{R}^{(n_x \times n_h)} \) and \( \phi_2 \in \mathbb{R}^{(n_u \times n_h)} \) are the high dimension feature space. Substituting Eq.~\ref{eq:lpv_3} into Eq.~\ref{eq:lpv_2} results in
\begin{equation}\label{lpvmain1}
\begin{split}
    \mathbf{x}({k+1}) &= \underbrace{\left[W_1\ \ W_2 \right]}_\text{\(\mathbf{W}\)}  \ 
    \underbrace{\left[
  \begin{array}{c}
\left(\phi_1\left(p(k)\right) \mathbf{x}(k)\right)^T \\
\left(\phi_2\left(p(k)\right) \mathbf{u}(k) \right)^T \\
  \end{array}
\right]}_\text{\(\Phi(k)^T\)}
\end{split}
\end{equation}
The residual error of modeling, $e(k) \in \mathbb{R}^{n_x} $, is defined as
\begin{equation}
\begin{split}
   e(k) = {W} \Phi(k)^T - \mathbf{x}({k+1})
\end{split}
\end{equation}
and it can be added to Eq.~\ref{lpvmain1} as
\begin{equation}\label{eq:res}
\begin{split}
   \mathbf{x}({k+1}) = {W} \Phi(k)^T +  e(k) 
\end{split}
\end{equation}
The LS-SVM cost function is then defined as
\begin{equation}\label{eq:optcov}
\begin{split}
\text{Minimize: } \ &\frac{1}{2} ||{W}||^2_2 + \frac{1}{2} \sum_{j=1}^N e(j)^T \gamma e(j) \\
\text{Subject to: \ } &
      \mathbf{x}({j+1}) = {W} \Phi(j)^T +  e(j) \\
  \end{split}
\end{equation}
Where \( N\) is the number of training samples used for modeling and \(j\) is the discrete sample time defined from 1 to \(N\). In this LS-SVM formulation, \( \gamma\) is a diagonal matrix of size \(n_x\) that acts as the regularization parameters. The Lagrangian function can then be calculated based on Eq.~\ref{eq:optcov} as
\begin{equation}\label{l}
\begin{split}
   L({W}) &= \frac{1}{2} ||{W}||^2_2 + \frac{1}{2} \sum_{j=1}^N e(j)^T \gamma e(j) \\ &- \sum_{j=1}^N \alpha_j^T \left( {W} \Phi(j)^T +  e(j) - \mathbf{x}({j+1}) \right)
\end{split}
\end{equation}
where \( \alpha_j^T \in \mathbb{R}^{n_x} \) are the discrete time Lagrange multipliers. To find the optimum \( W\), the derivatives of the Lagrangian, Eq.~\ref{l}, with respect to optimization variables must be zero as
 \begin{subequations}
\begin{flalign}
    \frac{\partial L}{\partial {W}} = 0& ~\rightarrow~ W = \sum_{j = 1}^N \alpha_j  \Phi(j)  \label{eq:ptod1}\\
    \frac{\partial L}{\partial e} = 0& ~\rightarrow~ \alpha = \gamma e  \label{eq:ptod2}\\
    \frac{\partial L}{\partial \alpha} = 0& ~\rightarrow~  {x}({j+1}) = {W} \Phi(j)^T +  e(j)  \label{eq:ptod3}
\end{flalign}
\end{subequations}
Substituting Eqs.~\ref{eq:ptod1} and \ref{eq:ptod3} into Eq.~\ref{eq:res} results in 
\begin{equation}\label{optres1}
\begin{split}
   {x}({k+1}) = \sum_{j = 1}^N \alpha_j \underbrace{\Phi(j) \Phi(k)^T}_{[\Omega]} +  \gamma^{-1} \alpha(k)
\end{split}
\end{equation}
where \(  \Phi(j) \Phi(k)^T \) is the kernel matrix, $[\Omega]$, and can be defined as
\begin{equation}\label{kernelmatrix}
\begin{split}
  [\Omega] &= {x}(j)^T K\left(p(j),p(k)\right){x}(k) \\
  &~~+ {u}(j)^T K\left(p(j),p(k)\right){u}(k)
\end{split}
\end{equation}
% \begin{equation}\label{kernelmatrix}
% \begin{split}
%   {\Omega} &= \left[\Phi(j) \ \ \ \Phi(k)^T\right] \\
%   &= \left[{x}(j)^T \underbrace{K\left(p(j),p(k)\right)}_{\text{Kernel function}}{x}(k) + {u}(j)^T \underbrace{K\left(p(j),p(k)\right)}_{\text{Kernel function}} {u}(k) \right]
% \end{split}
% \end{equation}
where $K(p(j),p(k))$ is a nonlinear kernel function. Usually, a Radial Basis Function (RBF) kernel, $K_{RBF}$, is used as the kernel function, which is defined as 
\begin{equation}
\begin{split}
    K_{RBF}\left(p(j),p(k)\right) = \exp\left(-\frac{||p(j)-p(k)||^2}{2\sigma}\right)
\end{split}
\end{equation}
where $\sigma$ is a free parameter that is tuned during the hyperparamter optimization and $||p(j)-p(k)||^2$ is the L$_2$ norm between the two feature vectors. Writing Eq.~\ref{optres1} in a compact notation yields
\begin{equation}
\begin{split}
   {X} = {\alpha} {\Omega} +  {\gamma}^{-1} {\alpha}
\end{split}
\end{equation}
where \( {X} = [x(1) \ \ ... \ \ x(N)] \). Solving this equation for \( {\alpha}\) results in
\begin{equation}\label{optres3}
\begin{split}
{\alpha} &= \left( I_N \odot \gamma^{-1} + \Omega^T \odot I_{n_x}\right)^{-1} {X}
\end{split}
\end{equation}
where \( I_N\) and \( I_{n_x}\) indicate the identity matrix in the dimension of training sample size by \( \mathbf{x}\) size and \( \odot\) is element-wise or Kronecker product. By calculating \( \alpha \), the state-space model matrices can be calculated as
\begin{equation}
\begin{split}
    A\left(p(k)\right) &= \sum_{j = 1}^N \alpha_j \mathbf{x}(j)^T K_{RBF}\left(p(j),p(k)\right)\\
    B\left(p(k)\right) &= \sum_{j = 1}^N \alpha_j \mathbf{u}(j)^T K_{RBF}\left(p(j),p(k)\right)\\
\end{split}
\end{equation}
where the $j$ index shows the data used in the training set. The model is developed using the training set of ${x}(j)~~~j \in (1,2,...,N)$ and ${u}(j)~~~j \in (1,2,...,N)$. Additionally, the scheduling parameter, $p$ is also given in the training set as $p(j)~~~j \in (1,2,...,N)$. 

\subsection{Bayesian Hyperparameters Optimization}

\begin{figure}[ht!]
    \centering
    \includegraphics[width = 0.49\textwidth]{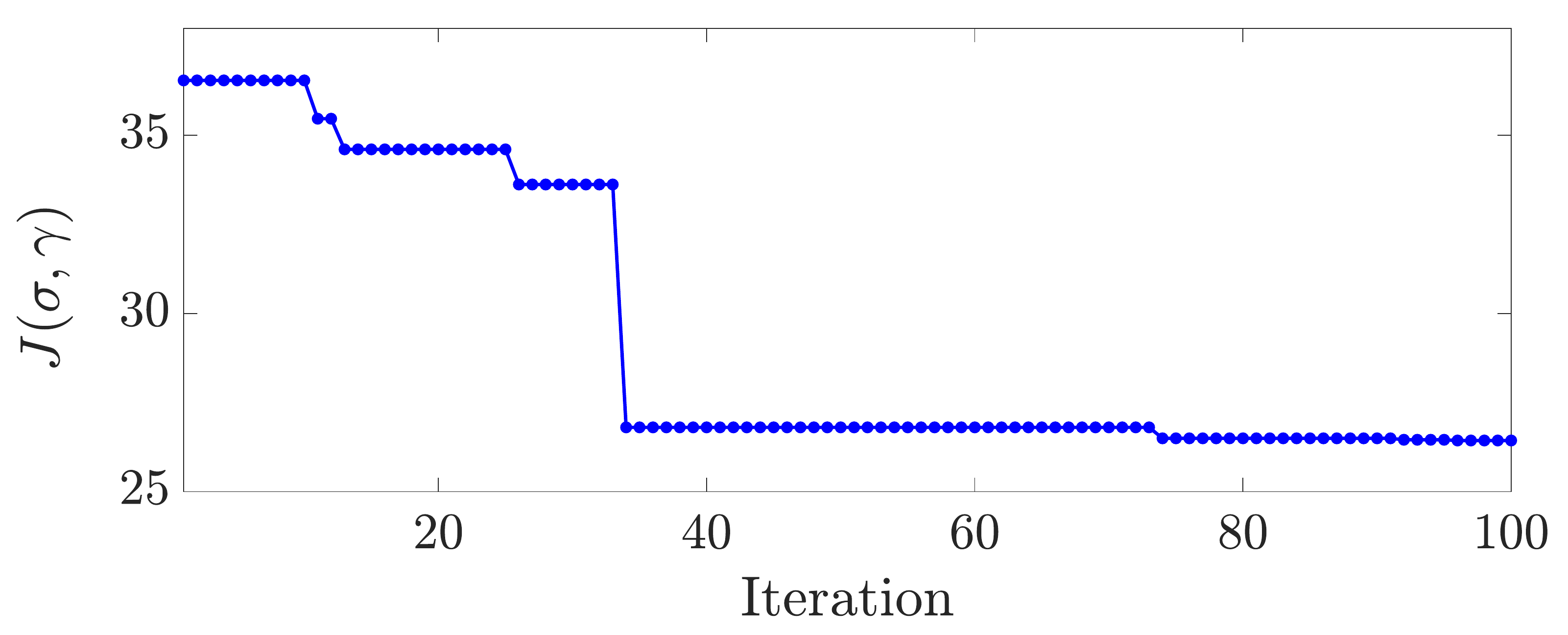}
    \caption{Bayesian optimization results for LPV-SVM model parameter optimization showing the cost function ($J$) values versus the integration number}
    \label{fig:lpvmodelbeys}
\end{figure}

\begin{figure}[ht!]
    \centering
    \includegraphics[trim = 0 50 40 40, clip, width = 0.48\textwidth]{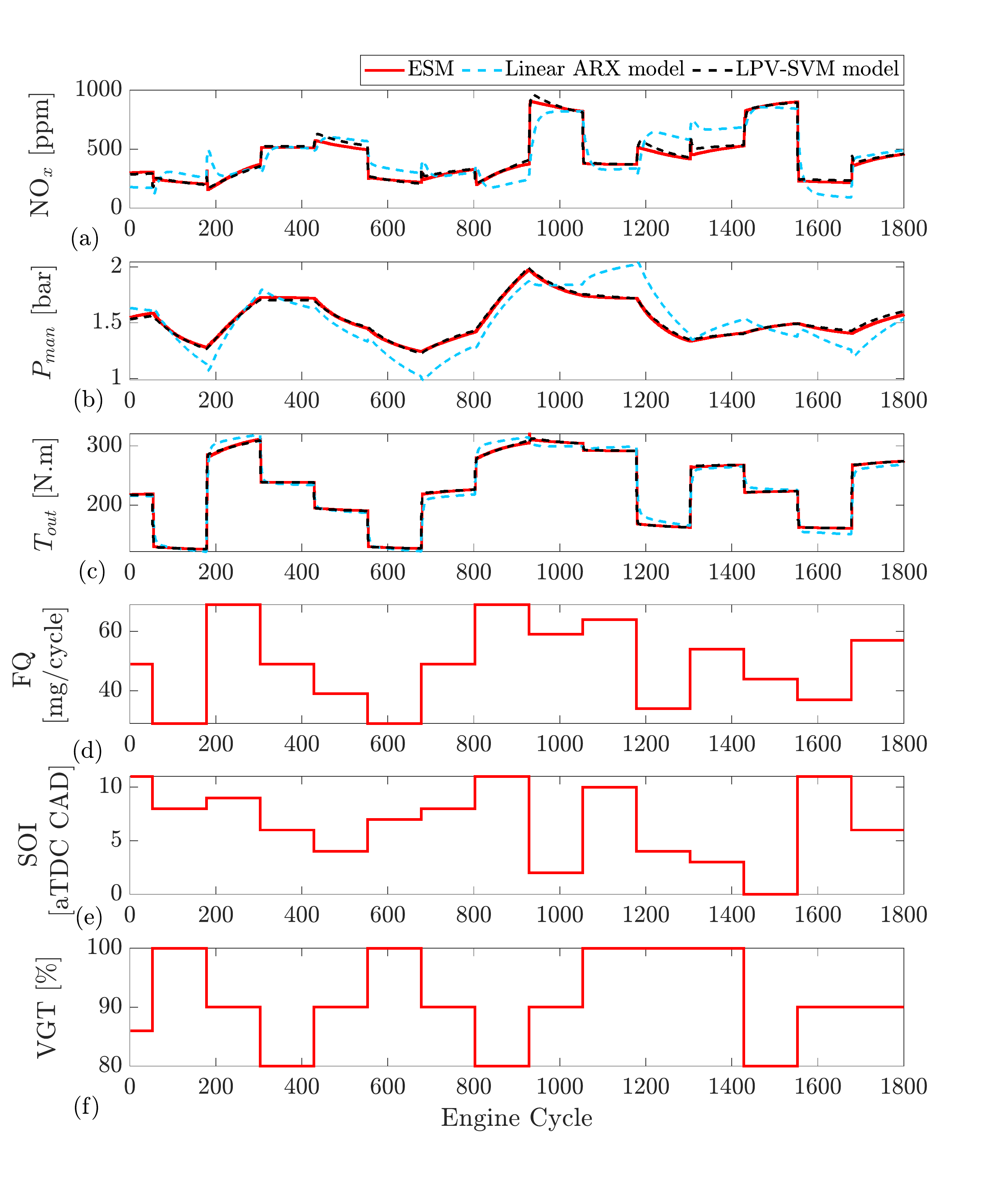}
    \caption{Linear ARX, LPV-SVM and ESM comparison for engine-out emissions and performance using ESM co-simulation: a) engine-out \nox, b) intake manifold pressure ($P_{\text{man}}$), c) engine output torque ($T_{\text{out}}$), d) Fuel quantity ($FQ$), e) Start of injection ($SOI$), f) Variable Geometry Turbine ($VGT$) rate}
    \label{fig:comparisonmodel}
\end{figure}

The SVM-LPV model has two main hyperparameters: \( \gamma\), the regularization coefficient, and \( \sigma\) the free parameters of the kernel. The cost function of the hyperparameter optimization is defined as 
\begin{equation}
    J(\gamma, \sigma) = \frac{1}{N_{CV}} \sum_{i = 1}^{N_{CV}} \left({\hat{X}}({i}) - {X}({i})\right)^2
\end{equation}
where \( \mathbf{\hat{x}}({i})\) is the modeled output and \(\mathbf{x}({i}) \) is the measured states and \(N_{CV}\) is the validation dataset that is used for optimizing the parameters. Bayesian Optimization utilizes Bayes Theorem to direct a search of a global optimization problem. The cost function versus iteration number for 100 iterations of the Bayesian optimization is shown in Fig.~\ref{fig:lpvmodelbeys}. In this figure, the Bayesian optimization approaches to the global optimum after 76 iterations. 
The Bayesian-optimized SVM-LPV can capture all states with an accuracy of 7.3\%, 1.1\%, and 1.9\% for \nox, \(T_{\text{out}}\), and \( P_{\text{man}}\) respectively when using the training data. As this model will be used for the Model Predictive Controller (MPC), the accuracy of this model for a new data-set is critical. The SVM-LPV is compared to a linear state-space model called Autoregressive with Extra Input (ARX) that is commonly used for system identification. The ARX-based discrete-time state-space model of the Diesel engine emissions and performance is trained using the same training dataset as the SVM-LPV resulting in: 
\begin{equation} \label{eq:linearsystem4}
\begin{split}
   A & = \left[
  \begin{array}{ccc}
0.7286 & 7.1252 & -0.0019  \\
0.0002 & 0.9859 & 8.9878 \times 10^{-6} \\
-0.6105 & 33.94287 & 0.9076 \\
  \end{array}
\right] \\
   B & = \left[
    \begin{array}{ccc}
1.2639 & -1.0899 & 1.0084 \times 10^{-5} \\
-0.0007 & 0.0014 & -1.01397 \times 10^{-5} \\
2.9360 & -8.2453 & -0.0106 \\
  \end{array}
\right] \\
     C & = \left[
  \begin{array}{ccc}
1 & 0 & 0  \\
0 & 0 &  1 \\
  \end{array}
\right]
\end{split}
\end{equation}
The linear ARX and SVM-LVP model are both run simultaneously in one simulation where both models are compared against the ESM. Both models are evaluated using unseen test data. Fig.~\ref{fig:comparisonmodel} shows the model comparison where both models have a high accuracy within 5\% normalized root mean square error (NRMSE) for estimating the output torque. However, the linear model fails to provide an accurate estimation for intake manifold pressure and \nox~ emissions. For the intake manifold pressure, the SVM-LPV model has significantly better estimation than the linear model where the NRMSE is 0.95\% in comparison to 14.81\% for the linear model. For \nox, the SVM-LPV estimates with less than 7\% error. While the linear ARX model is unable to accurately capture the emission level resulting in a NRMSE of 32.3\%. Next, using the developed SVM-LPV model, an MPC combustion controller will be designed. 

% \begin{table}
%     \centering
%      \caption{Comparison of linear and LPV models for new generated test data}
%     \begin{tabular}{c | c c c}
%     \hline
%         & \textbf{NO}$_x$  & \textbf{\(T_{\text{out}}\)} & \textbf{\(P_{\text{man}}\)} \\
%         \hline
%         \textbf{Linear} & 32.3\% & 4.04\% & 14.81\% \\
%         \textbf{SVM-LPV} & 6.95\% & 3.02\%  & 0.96\%\\
%         \hline
%     \end{tabular}
%     \label{tab:comparisonmodel}
% \end{table}

% Here 3750 engine cycles are used for the training of the ARX model. The error between the ARX model and GT-Power virtual plant results in a normalized root mean square error (NRMSE) of 15.25\%, 18.19\%, 3.85\% for \nox~ emissions, intake manifold pressure (\(P_{\text{man}}\)) and output torque (\(T_{\text{out}}\)), respectively. This model will used for sake of comparison between LPV model for a test data set that was not used for training will be presented. 

%%%%%%%%%%%%%%%%%%%%%%%%%%%%%%%%%%%%%%%%%%%
\section{Model Predictive Controller Design}
\subsection{Controller Design}

The controller's objective in this study is to minimize engine-out \nox~emissions and fuel consumption while maintaining the desired engine output torque. The cost function $J (\bm{u}(\cdot|k))$ of the finite horizon optimal control problem (OCP) with horizon length $ N_p $ is defined as
\begin{equation} \label{eq:mpccostfunc}
\begin{split}
    &J(\bm{u}(\cdot|k), s(k)) =\\ &\sum_{i = 0}^{N_p-1} \Big[ \underbrace{||T_{\text{out}}(k+i) - T_{\text{out, ref}}(k+i) ||^2_{w_{T_{\text{out}}} }}_{\text{Torque output tracking}} \\
    &+ \underbrace{||\text{NO}_x(k+i)||^2_{w_{\text{NO}_x}}}_{\text{NO$_x$ minimizing}} + \underbrace{||FQ(k+i)||^2_{w_{FQ}}}_{\text{fuel consumption minimizing}}\\
    &+ \underbrace{||u(k+i|k) - u (k +i -1|k)||^2_{w_{\Delta u}}}_{\text{control effort penalty}} \\
    &+ \underbrace{w_{s} s(k)^2 }_{\text{Constraint violation penalty}} \Big]
\end{split}
\end{equation}

where 
\begin{equation}
    ||.||^2_w = [.]^Tw[.]
\end{equation}

where $s(k)$ is a slack variable that is added to the cost function to allow slight violation of the constraints. In this equation, $w_{v}$, $ v \in [T_{\text{out}}, \text{NO}_x, FQ, \Delta u, s ]$ are the MPC weights. The optimization decision, $\bm{u}(\cdot|k)$, is defined as
\begin{equation}
    \bm{u}(\cdot|k) = [ u(k|k)^T \ \   u(k+1|k)^T \ \ ... \ \ u(k+N_p-1|k)^T]
\end{equation}

Based on the defined cost function, the Optimal Control Problem (OCP) solved at each discrete-time instance, (an engine cycle) is
\begin{mini!}|s|
	{\substack{\bm{u}(\cdot|k), \\s(k)}} %, s_0,\ldots, s_{N}
	{J\left(\bm{u}(\cdot|k), s(k)\right)} %\oj{+ \sum_{i=0}^{N} \norm{s_k}}
	{\label{eq:OCP}}
	{}
	\addConstraint{x({0})}{ = \bar{x}(0)}
	\addConstraint{x({k+1})}{=f(x(k),u(k)) ~~}{\forall  k\in[0, N_p-1]}
	\addConstraint{\underline{x}}{\leq x_k \leq \overline{x}}{\forall  k\in[0, N_p]}
	\addConstraint{\underline{u}}{\leq u_k \leq \overline{u}}{\forall  k\in[0, N_p-1]}
\end{mini!}
Where $f(x(k),u(k))$ are based on Eq.~\ref{eq:lpv_2} which is the SVM-LPV model of the system and $x$ are the states of model. A five step prediction horizon is used due to the dynamics of the \nox~emissions and a single step control horizon is used. For the linear MPC, $A$ and $B$ are constant matrices. The optimization is subject to the constraints listed in Table~\ref{tab:mpc_constraints}. The 500 ppm maximum \nox~ is chosen as the upper limit to keep the emissions below the maximum experimentally measured \nox~output from the Cummins calibrated engine controller. However, this constraint can be adjusted depending on emission legislation. The limit on FQ is used as a safety constraint. Limits in SOI are added to avoid early combustion which can lead to combustion noise and late combustion which can lead to low thermal efficiency and high exhaust gas temperatures. The turbocharger characteristic map is used to set the VGT limit.

\begin{table}[hb!]
\begin{center}
\caption{Constraint Values}\label{tab:mpc_constraints}
\begin{tabular}{ccc}
\hline
Min Value ($\underline{x}, \underline{u})$ & Variable $(x,u)$ & Max Value $(\overline{x}, \overline{u})$ \\ \hline
$0\ \mathrm{ppm}$ & $NO\ind{x}$ & $500\ \mathrm{ppm}$ \\
$10\ \mathrm{mg/cycle}$ & $FQ$ & $80\ \mathrm{mg/cycle}$ \\
$-2\ \mathrm{aTDC~CAD}$& $SOI$ & $11\ \mathrm{aTDC~CAD}$ \\
$70\ \mathrm{\%}$& $VGT$ & $100\ \mathrm{\%}$\\
\hline
\end{tabular}
\end{center}
\end{table}

% For the linear MPC, the model dynamics of the discrete-time-state-space model given in \eqref{eq:linearsystem2} are utilized as $x_{k+1}= f(x_k, u_k)$. The linear MPC formulation has been implemented in MATLAB using the MATLAB Linear MPC block, \texttt{mpc(sys)} and the corresponding linear MPC block in Simulink. 

% For the LPV model, the model matrices are changed based on scheduling parameters identified using SVM techniques. Here the model dynamics are given in Eq.~\eqref{eq:lpv_2} which are utilized as $x_{k+1}= f(x_k, u_k)$. For this, the MPC structure is defined in the MATLAB MPC toolbox, then using the MATLAB Adaptive MPC block the system matrices $A$ and $B$ are updated using the defined scheduling parameters. For both the linear and LPV MPC, Mathwork fmincon function has been used. In both of these models the state vector is defined as
% \begin{align}
% \begin{split}
%     x(k) &= \begin{bmatrix} T_\text{out}(k) & p_{\text{\text{man}}} & \text{NO}_x(k) \end{bmatrix}^T
% \end{split}
% \end{align}

% The weights of the Linear and LPV MPC values are $w_{T_{\text{out}}} = 0.009$, $w_{\text{NO}_x} = 0.0004$, $w_{FQ} = 0.06 $, $w_{\Delta u} = 0.1$. The constraint softening value was set as 0.1 which indicates hard constraints. 
 
\subsection{Controller Results}

\begin{figure}[ht!]
     \centering
    \includegraphics[trim = 0 50 40 35, clip,width = 0.48\textwidth]{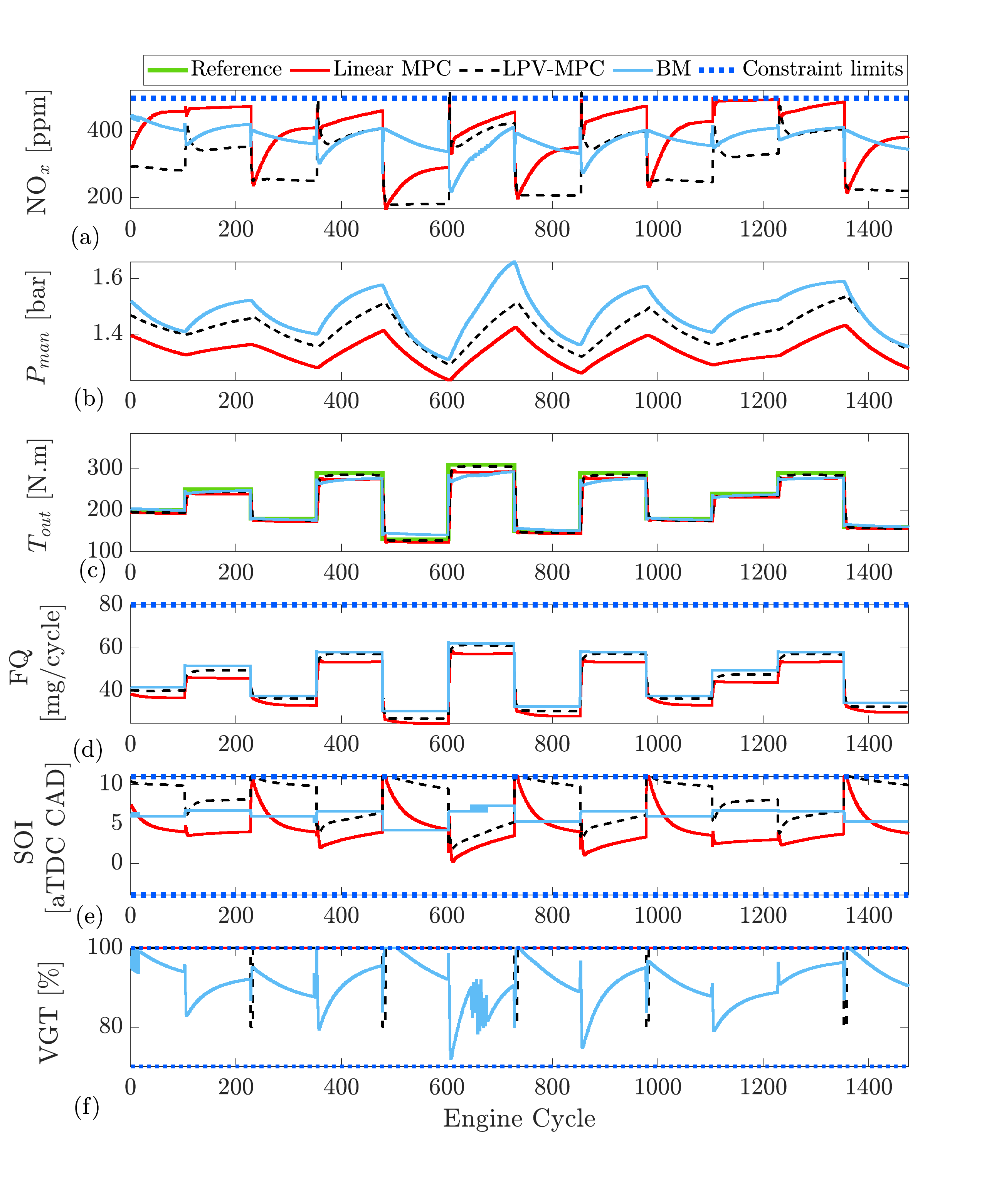}
    \caption{Linear MPC, LPV-MPC and Benchmark comparison at 1500: a) engine-out \nox, b) intake manifold pressure ($P_{\text{man}}$), c) engine output torque ($T_{\text{out}}$), d) fuel quantity ($FQ$), e) Start of injection ($SOI$), f) Variable Geometry Turbine ($VGT$) rate}
    \label{fig:com1500}
\end{figure}

As the LPV model is developed at a constant speed of 1500 rpm, the ESM is also simulated at that speed. The developed MPC based on the SVM-LPV model, the linear ARX-based linear MPC and benchmark (BM) ESM calibrated ECU based on the Cummins production ECU are compared and shown in Fig.~\ref{fig:com1500}. Except for a slight violation in \nox~constraints (for example at engine cycle 600), both controllers are able to keep \nox~emissions below the specified constraint. For the \nox~emissions, both the linear MPC and BM controller have higher overall emissions levels than the LPV-MPC controller. This is likely due to the simple linear model used in the linear MPC that is unable to capture the non-linear \nox~formation trends. 

As shown, both the LPV-MPC and Linear MPC (LMPC) tend to generate late injection timings which cause a reduction in peak combustion temperature resulting in lower \nox~levels. However, this late combustion phasing can result in lower thermal efficiency and higher fuel consumption. For this reason an upper bound is added for SOI. The output torque ($T_{\text{out}}$) tracking performance is within 5\% for all three controllers. The feed-forward controller does fail to reach the target torque and remains slightly below the set-point for each step.

\begin{figure}[h!]
    \centering
    \includegraphics[trim = 0 0 0 0, clip, width = 0.48\textwidth]{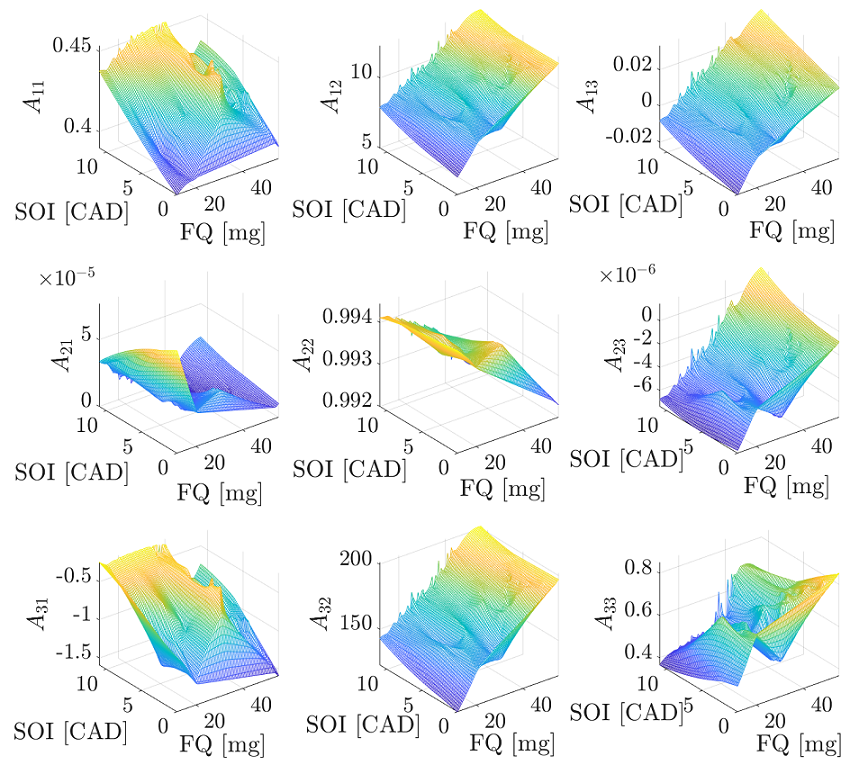}
    \caption{``A" matrix elements for the LPV-SVM model at the engine speed of 1500 rpm}
    \label{fig:a1500}
\end{figure}

The LPV-SVM model contains a gain scheduling matrix $A$ and $B$ which are dependent on the inputs SOI and FQ.  The scheduling parameters as a function of inputs for matrix $A$ are shown in Fig.~\ref{fig:a1500} which shows that the relationship between the model inputs and the scheduling parameters are non-linear for the Diesel combustion process. This non-linearity of the gain scheduling variables of the LPV-SVM model are an advantage of using the LPV model for the combustion instead of using only a few points of linearization. The gain scheduling matrix B is similarly nonlinear. 

% \begin{figure}[h!]
%     \centering
%     \includegraphics[trim = 0 0 0 0, clip, width = 0.48\textwidth]{Figures/B.eps}
%     \caption{A matrix for LPV-SVM model in engine speed of 1500 rpm}
%     \label{fig:b1500}
% \end{figure}

\begin{figure}
     \centering
     \centering
    \includegraphics[trim = 0 50 40 40, clip,width = 0.46\textwidth]{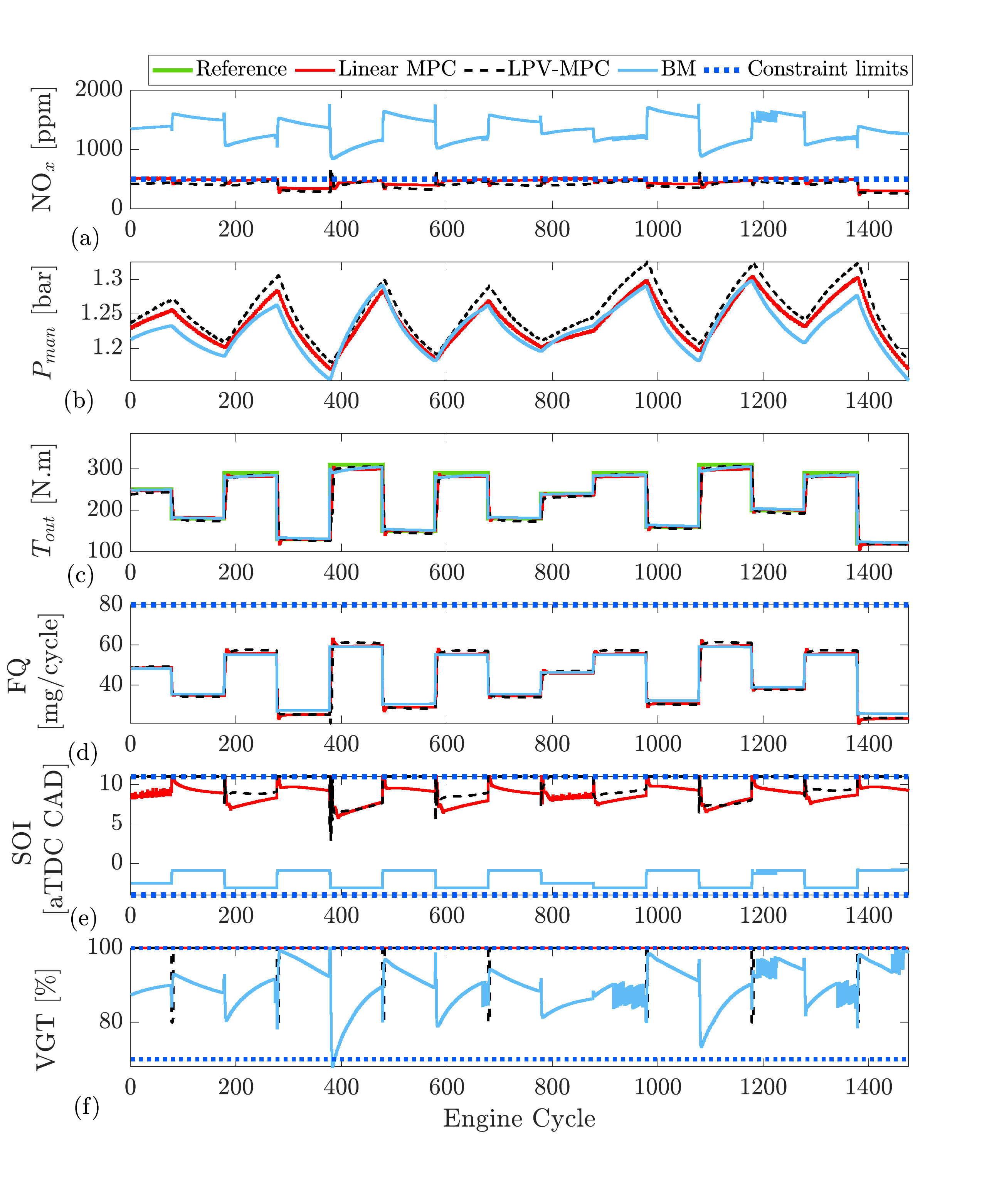}    \caption{Linear MPC, LPV-MPC and Benchmark comparison at 1200 rpm: a) engine-out \nox, b) intake manifold pressure ($P_{\text{man}}$), c) engine output torque ($T_{\text{out}}$), d) Fuel quantity ($FQ$), e) Start of injection ($SOI$), f) Variable Geometry Turbine ($VGT$) rate}
    \label{fig:com1200}
\end{figure}

The LPV model is developed at a constant speed of 1500 rpm. To evaluate the controller robustness, each controller is tested at an engine speed of 1200 rpm. As shown in Fig.~\ref{fig:com1200}, both the LMPC and LPV-MPC perform significantly better than BM. Here the benchmark controller tends to advance injection timing at lower speeds which results in significant increases in \nox~emissions. Due the increased accuracy of the LPV model, LPV-MPC performs slightly better. In the next section MPC will be replaced by imitation ML to reduce computation.

%%%%%%%%%%%%%%%%%%%%%%%%%%%%%%%%%%%%%%%%%%%
\section{Imitation of MPC using a Deep Neural Network}

Machine learning was used to model the system in section 5. Here ML is used to replace the MPC with a learning controller called imitative LPV-MPC. The goal is to avoid the high computational time of MPC, that requires solving MPC optimization online. Instead, a function, in this case a deep network, is trained to approximate the MPC and can be deployed with a much lower computational cost. 

\begin{figure}[h!]
    \centering
    \includegraphics[trim = 0 0 0 0, clip, width = 0.45\textwidth]{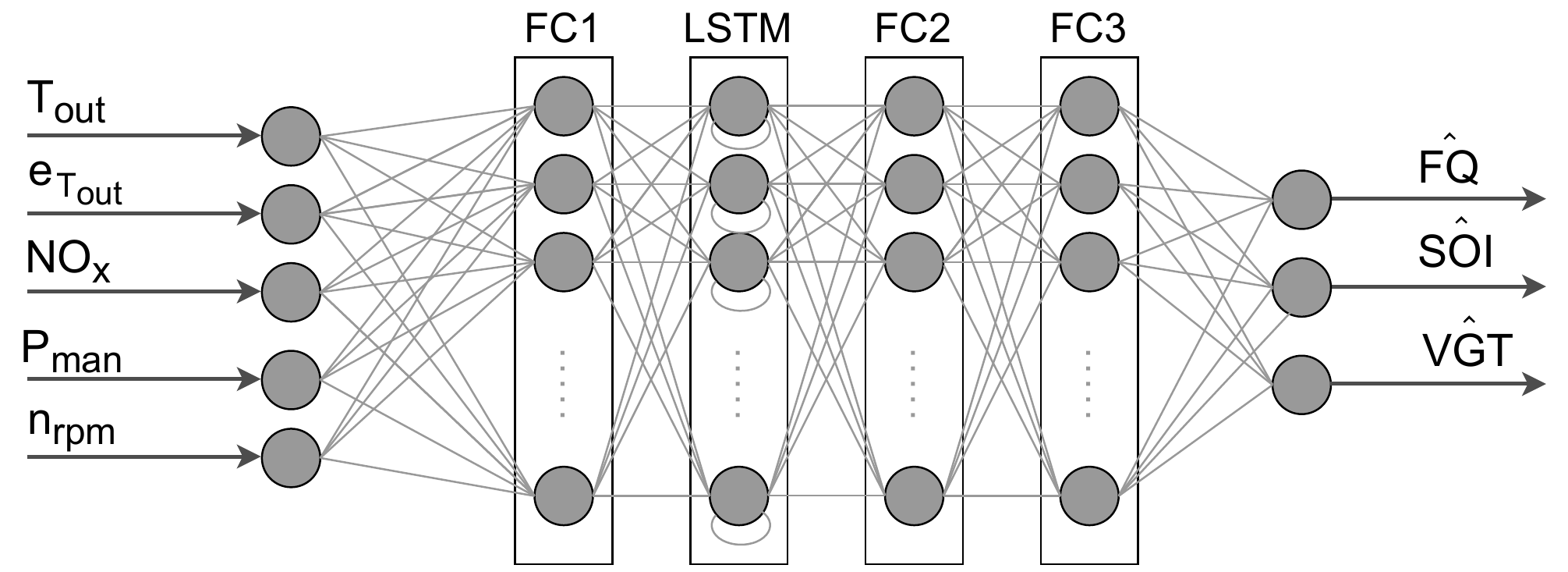}
    \caption{Structure of proposed network for imitation of LPV-MPC}
    \label{fig:StructureOfimitative}
\end{figure}

The schematic of imitative LPV-MPC was previously shown in Section 4 of Fig.~\ref{fig:offlinemodeling} (step 4). First, the LPV-MPC are implemented using the ESM in simulation. Then the controller input and outputs are recorded, and a deep neural network, including a Long-Short-Term Memory (LSTM) layer, are used to mimic the behavior of the MPC as shown schematically in Fig.~\ref{fig:StructureOfimitative}. The inputs of this network are engine output torque, the error in output torque (\( e_{T_{\text{out}}}\)), engine-out \nox, intake manifold pressure \( P_{\text{man}} \), and engine speed \( n_{rpm}\) and the outputs are fuel quantity \( (FQ) \), start of injection \( (SOI) \), and VGT. This network includes four main layers where the first, third and fourth layer are fully connected (FC) layers with a layer size (neurons) of 32. The second layer is an LSTM layer with the same layer size. The reason for using an FC layer around the LSTM to create a deep network is to increase the complexity of the model without increasing the number of hidden and cell states of the entire network. Finally, the online MPC is replaced with the designed imitative LPV-MPC. 

\begin{figure}[ht!]
     \centering
     \centering
    \includegraphics[trim = 0 50 40 40, clip,width = 0.46\textwidth]{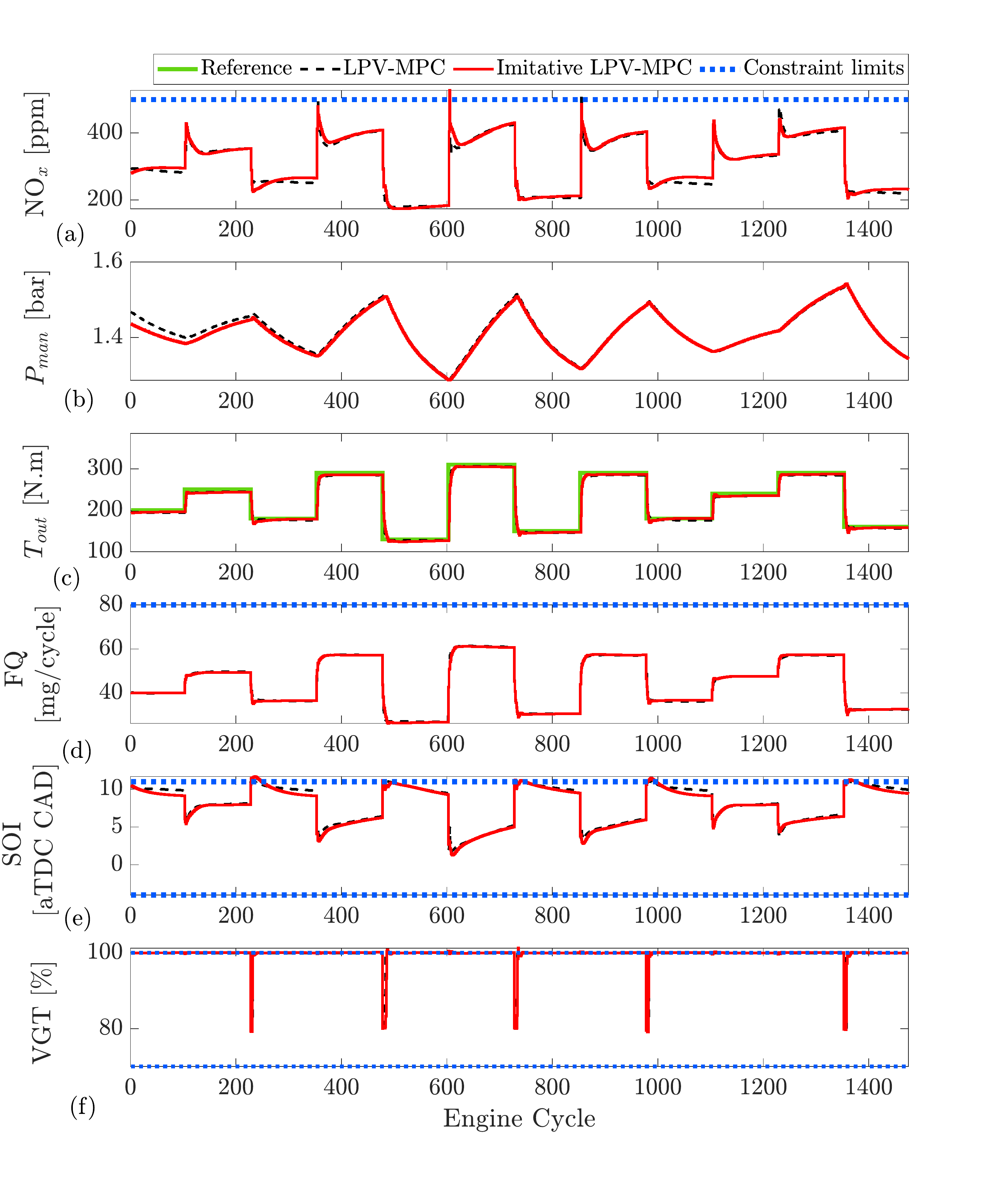}
    \caption{LPV-MPC and imitative LPV-MPC comparison in 1500 rpm: : a) engine-out \nox, b) intake manifold pressure ($P_{\text{man}}$), c) engine output torque ($T_{\text{out}}$), d) Fuel quantity ($FQ$), e) Start of injection ($SOI$), f) Variable Geometry Turbine ($VGT$) rate}
    \label{fig:imitative1500}
\end{figure}

To train the imitative controller network, the LPV-MPC is evaluated for 2000 seconds at different engine speeds with randomly generated output torque targets. Of these 2000 seconds, 1600 seconds are devoted to training and the remainder to validation. To train this network, a mini-batch size of 512, initial learn rate of 0.01, and a learn rate drop period of 200 Epochs with a drop factor of 0.5 is used. In this training, an L$_2$ Regularization with a value of 0.8 is used. The training accuracy for FQ, SOI, and VGT is 4.3\%, 6.3\%, and 8.3\% while for the validation data an accuracy of 4.3\%, 8.9\%, 10.3\% is observed. To test the imitative LPV-MPC, the controller is tested on a previously unseen reference and compared to the LPV-MPC in Fig.~\ref{fig:imitative1500}. The results show that the imitative controller can successfully clone the behaviour of the LPV-MPC and generate approximately optimal control without performing an online optimization.

% \begin{figure}[ht!]
%      \centering
%      \centering
%     \includegraphics[trim = 10 50 40 40, clip,width = 0.46\textwidth]{Figures/imitation_vs_lpv_1200rpm.eps}    \caption{LPV-MPC and imitative LPV-MPC comparison in 1200 rpm:a) engine-out \nox, b) intake manifold pressure ($P_{\text{man}}$), c) engine output torque ($T_{\text{out}}$), d) Fuel quantity ($FQ$), e) Start of injection ($SOI$), f) Variable Geometry Turbine ($VGT$) rate}
%         \label{fig:imitative1200}
% \end{figure}

The performance of the controllers at engine speeds of 1500 and 1200 rpm are compared to the baseline model and are summarized in Table~\ref{tab:savingcontroller}. As the LPV controller was designed only based on constant speed data at 1500 rpm, the LPV-MPC and imitative controller's performance show the controllers robustness changing speeds. Here a significant \nox~emissions reduction can be seen for all the controllers over the baseline model except for the LMPC model at 1500~rpm which can be attributed to the use of a simplified linear model. In addition to the reduced emissions for all controllers, they are able to maintain or improve fuel consumption compared to the baseline. This demonstrates the advantage of the optimized controllers over the calibration based baseline. One disadvantage of the developed models is an increase in load tracking error in comparison to the baseline model at 1500~rpm. However, this 2\% discrepancy in load tracking results in significant emission reduction of 18-70\% and fuel consumption reduction of 1-10\%.

\begin{table*}[ht!]
    \centering
    \caption{Percentage of improvement for proposed MPC and imitative LPV-MPC with respect to the Benchmark for engine speeds of 1500 and 1200 rpm}
    \begin{tabular}{l | c c c c}
        \hline
        \hline
        \multicolumn{5}{c}{\textbf{1500 rpm}}\\
        \hline
        \hline
         & \nox~ [\%]  &  FQ [\%] & load error [\%] & Average time per cycle [ms] \\
        \hline
        LMPC & +6.71 & -10.00 &  -1.04  & 1.17\\
        LPV-MPC & -18.98 & -3.48 & +0.99 & 1.69\\
        Imitative LPV-MPV & -18.05 & -3.70 & +1.03 & 0.03\\
        \hline
        \hline
        \multicolumn{5}{c}{\textbf{1200 rpm}}\\
        \hline
        \hline
        & \nox~ [\%]  & FQ [\%] & load error [\%] & time per cycle [ms] \\
        \hline
        LMPC & -66.20 & -1.23 & -0.98 & 1.93\\
        LPV-MPC & -69.77 & 0.00 & -0.88 & 1.61\\
        Imitative LPV-MPV  & -70.48 & -1.85 & -3.56 & 0.03\\
        \hline
    \end{tabular}
    \label{tab:savingcontroller}
\end{table*}

The imitative LPV-MPC controllers provide similar improvements to the full MPC controllers over the baseline model, while providing significantly improved computational times. As presented in Table~\ref{tab:savingcontroller}, the imitative controllers are 50 and 77 times faster than online MPC optimization at 1500 and 1200 rpm, respectively. All these simulations are carried out on a computer equipped with Intel Core i7-6700K processor with 32.0 GB of RAM. This computation requirement with respect to online MPC optimization makes future real-time implementation of this controller feasible. 

\section{Summary and Conclusions}

This paper presents the integration of machine learning and model predictive control for both modeling and controller implementation for a diesel engine application. First, a support vector machine based linear parameter varying model is developed to design an LPV-MPC. When comparing the results of linear and LPV models for newly generated inputs, the LPV model showed better prediction accuracy for all engine outputs. Using these models a linear MPC and an LPV-MPC are designed. 

Then, the LPV-MPC is implemented and the controller input and output data are collected from the MPC and used to train a deep neural network. By replacing the full online MPC with a deep network the aim is to reduce the computational time of the MPC. After testing the imitative LPV-MPC controller at two different engine speeds, the imitative controller performs very closely to the online optimized MPC but with a significant reduction in the processing time. In addition, the MPC and imitative models showed significant improvements in \nox~emissions and a reduction in fuel consumption while providing similar load following capabilities as the feed-forward production controller. Both the LPV-MPC and imitative controller are able to reduce \nox~emissions by 18-70\% while reducing fuel consumption by 1-10\% compared to the Cummins production controller and the imitative controller requires 1/50 the computation time compared to online MPC optimization.

\section*{Acknowledgments}
The author(s) disclosed receipt of the following financial support for the research, authorship, and/or publication of this article: The research was performed as part of the Research Group (Forschungsgruppe) FOR 2401 “Optimization based Multiscale Control for Low Temperature Combustion Engines,” which is funded by the German Research Association (Deutsche Forschungsgemeinschaft, DFG) and with Natural Sciences Research Council of Canada Grant 2022-03411. Partial funding from Future Energy Systems at the University of Alberta is also gratefully acknowledged.

\bibliography{ifacconf}

% \appendix

\end{document}